\documentclass[runningheads]{llncs}
\usepackage[T1]{fontenc}
\usepackage{graphicx}
\usepackage{xcolor}
\usepackage{hyperref}
\hypersetup{
    colorlinks=true,
    linkcolor=blue,
    citecolor=blue,
    filecolor=blue,
    urlcolor=blue
}
\usepackage{tabulary}
\begin{document}
\title{User Concerns Regarding Social Robots for Mood Regulation: A Case Study on the “Sunday Blues”}
\titlerunning{Social Robots for Mood Regulation}
%
\author{Zhuochao Peng\inst{1}\orcidID{0000-0001-7855-479X} \and
Jiaxin Xu\inst{2}\orcidID{0000-0003-3487-6281} \and
Jun Hu\inst{2}\orcidID{0000-0003-2714-6264} \and
Haian Xue\inst{3}\orcidID{0000-0002-2351-9070} \and
Laurens A. G. Kolks\inst{1}\orcidID{0000-0002-5636-2495} \and
Pieter M. A. Desmet\inst{1}\orcidID{0000-0002-0244-5359}}
\authorrunning{Z. Peng et al.}
%
\institute{Delft University of Technology, Delft, the Netherlands\\
\email{\{z.peng, l.a.g.kolks, p.m.a.desmet\}@tudelft.nl} \and
Eindhoven University of Technology, Eindhoven, the Netherlands\\
\email{\{j.xu2, j.hu\}@tue.nl} \and
Tongji University, Shanghai, China\\
\email{\{haianxue\}@tongji.edu.cn}}
\maketitle              
\begin{abstract}
While recent research highlights the potential of social robots to support mood regulation, little is known about how prospective users view their integration into everyday life. To explore this, we conducted an exploratory case study that used a speculative robot concept—Mora—to provoke reflection and facilitate meaningful discussion about using social robots to manage subtle, day-to-day emotional experiences. We focused on the “Sunday Blues,” a common dip in mood that occurs at the end of the weekend, as a relatable context in which to explore individuals’ insights. Using a video prototype and a co-constructing stories method, we engaged 15 participants in imagining interactions with Mora and discussing their expectations, doubts, and concerns. The study surfaced a range of nuanced reflections around the attributes of social robots like empathy, intervention effectiveness, and ethical boundaries, which we translated into design considerations for future research and development in human-robot interaction.

\keywords{Design consideration \and Human-robot interaction \and Mental health and well-being \and Mood regulation \and Social robots.}
\end{abstract}
\section{Introduction}
The use of social robots to support mental health and well-being has been explored across domains. However, it is only in recent years that this topic has emerged as a clearly defined area in human-robot interaction (HRI)~\cite{ref_laban1}. In response to this growing interest, several recent systematic reviews have analyzed existing studies on the subject~\cite{ref_scoglio,ref_robinson,ref_nichol,ref_ghafurian}. These reviews suggest that most current research focuses on addressing severe emotional distress (e.g., depression or stress/anxiety disorders), primarily in therapeutic or clinical settings (e.g., mental healthcare facilities), or among vulnerable populations (e.g., children or older adults with/without dementia). In contrast, the idea that social robots could help manage everyday mood fluctuations—the subtle shifts in emotional state that many people experience as part of daily life—remains largely underexplored within the HRI community.

Human moods fluctuate, like the ever-changing weather—sometimes bright with happiness, other times clouded by sadness. Moods are low-intensity, diffuse feeling states that typically persist for hours~\cite{ref_morris1}. They are always present, gradually evolving, and often operating below conscious awareness~\cite{ref_watson}. Despite their subtle nature, moods directly influence subjective well-being. When people experience a positive mood, they tend to evaluate their overall life as more satisfying and fulfilling, and they are more likely to recall positive life events compared to when they are experiencing a negative mood~\cite{ref_morris2}. Furthermore, moods can significantly impact overall health. Persistent negative moods can contribute to mental health problems such as affective disorders~\cite{ref_peeters} and increase the risk of physical health issues like heart disease~\cite{ref_cohen}. Additionally, moods influence daily functioning and performance by affecting individuals’ perceptions, judgments, and decision-making~\cite{ref_forgas}. Given these profound effects mood has on individuals, effective mood regulation becomes essential in daily life~\cite{ref_larsen,ref_parkinson}.

Looking toward a future where social robots become part of people’s personal lives as close companions, they hold promise for supporting everyday mood regulation. These robots could encourage people to open up and express their feelings through empathetic, non-judgmental communication~\cite{ref_laban2}. They might also offer personalized, context-aware suggestions to promote self-care practices~\cite{ref_arango}. Compared to disembodied agents like chatbots, social robots can leverage additional communication channels, including proxemics~\cite{ref_takayama}, oculesics~\cite{ref_mutlu}, and physical contact~\cite{ref_willemse}. Utilizing these channels can enhance emotional exchanges and make interactions with robots feel more comforting~\cite{ref_bates}, engaging~\cite{ref_kidd}, and helpful~\cite{ref_fasola}. Additionally, the heightened social presence of robots can facilitate rapport building, providing not only immediate but also sustained company and support~\cite{ref_deng}.

Recent research in HRI has begun exploring the use of social robots for everyday mood regulation. For instance, Jeong et al.~\cite{ref_jeong} designed a social robot intervention for college students living in dormitories, where the robot helped with everyday tasks and engaged students through casual conversation. Their results indicated that interactions with the robot positively influenced students’ overall moods. Similarly, Laban et al.~\cite{ref_laban3} investigated the effects of self-disclosure to a social robot on caregivers, a group often experiencing distress. Their study found that participants who shared their thoughts and feelings with the robot reported improved moods and perceived the robot as increasingly comforting over time. Despite these promising findings, integrating social robots into daily life can raise significant concerns. Studies have identified such user concerns in various contexts. For instance, in early childhood education, teachers worry about their limited knowledge about robots, safety risks, and potential distractions for children~\cite{ref_neumann}. Similarly, implementing social robots for people with dementia faces challenges such as their unfamiliarity with technology, fear of robots, and privacy issues~\cite{ref_koh}. This suggests that while social robots have the potential to support everyday mood regulation, they may also introduce major problems that hinder user acceptance and robot employment. However, existing research on social robots for mood regulation has only briefly touched upon these user concerns (e.g.,~\cite{ref_axelsson}). Given the rapid advancements in social robotics, especially with large language models enabling more sophisticated social interactions, understanding users’ concerns is just as crucial, if not more so, than exploring their benefits.

Hence, this study aims to address a key research question: \textit{What concerns do users have about integrating social robots into their daily lives for mood regulation, particularly for managing everyday subtle mood fluctuations?} To achieve this aim, we conducted an exploratory case study investigating prospective users’ attitudes and opinions on using a social robot to manage the “Sunday Blues”—a common negative mood state experienced during the transition from the weekend to the workweek. Insights from this case study provide a foundation for understanding end-user expectations and concerns, offering design recommendations for future applications of social robots in mood regulation.
\section{Method}
\subsection{The Case Study}
We selected the Sunday Blues as the focus of our case study—a mood characterized by anxiety, sadness, or regret as the weekend concludes and the new workweek approaches~\cite{ref_zuzanek}. Its typical causes include the loss of leisure time, unmet weekend expectations, and anticipation of upcoming workloads and challenges~\cite{ref_tufvesson}. A recent survey suggests this mood issue is widespread among employees, with 80\% of respondents reporting frequent experiences of it~\cite{ref_heitmann}. Given its prevalence and impact on employees’ mental health and well-being~\cite{ref_akay,ref_mihalcea}, the Sunday Blues has gained significant attention in popular culture, with numerous blogs and podcasts addressing the topic and suggesting coping strategies (e.g.,~\cite{ref_headspace,ref_calm}). Despite this, the phenomenon remains largely unexplored in academic research, including within the HRI research community, highlighting an opportunity to explore potential solutions based on social robots. Leveraging the Sunday Blues as a relatable context, we designed a robot aimed at helping individuals manage or alleviate this negative mood.

To illustrate the robot’s functionality, we developed a video prototype depicting key interactions between users and the robot. We chose \textit{video prototyping} to elicit feedback on a concept that is not yet technically feasible in the form envisioned. This method allowed participants to immerse themselves in a realistic scenario and reflect on how the robot might fit into their own routines. Video-based scenarios are a common method in early-stage HRI research (e.g.,~\cite{ref_albers,ref_syrdal1,ref_xu}), particularly when exploring emotionally sensitive topics or future-use contexts, as they avoid the ethical and practical constraints of live deployment while still enabling rich user engagement~\cite{ref_zamfirescu-pereira,ref_syrdal2}.

To facilitate discussions with potential users, we employed the method of \textit{co-constructing stories}, engaging participants in direct dialogue to envision and articulate their thoughts about a novel design based on personal lived experiences~\cite{ref_buskermolen}. This method has proven helpful in eliciting in-depth user feedback and suggestions in various design contexts (e.g.,~\cite{ref_christiansen,ref_xue,ref_davis}). It is important to emphasize that the purpose of this study was not to evaluate a functional product, but to open up a design-led inquiry into the emerging space of social robots for everyday mood regulation. By presenting a speculative concept, we sought to provoke reflection and facilitate meaningful discussion with prospective users. Through these conversations, we explored how participants imagine living with a mood-regulating social robot, what expectations, hopes, and concerns this raises, and what these reflections teach us about the broader challenges of designing emotionally supportive technologies for everyday life.

In the following sections, we describe our design and video prototyping process, participant recruitment, co-constructing stories sessions, and data analysis.

\subsection{Design and Video Prototyping}
\subsubsection{Design Concept}
We conceptualized \textit{Mora} (short for “Mood Regulation Assistant”), a social robot designed to function both as an everyday companion and as a personal assistant within home environments. Mora’s primary goal is to monitor users’ mood fluctuations and provide timely emotional support. One key focus of Mora is helping users cope with the Sunday Blues during the transitional period from the weekend to the weekdays. Specifically, once detecting signs of anxiety, sadness, or unease emerging on Sunday evenings or nights, Mora approaches users and offers conversation-based interventions to alleviate their negative feelings. This concept is inspired by recent HRI research, which highlights the mood-regulatory benefits of sharing thoughts and feelings with a social robot (e.g.,~\cite{ref_laban3,ref_akiyoshi,ref_duan}). Mora’s intervention approach incorporates the following three evidence-based psychological strategies for mood regulation.

The first strategy is \textit{venting}~\cite{ref_zech}. To interrupt and prevent a potential emotional spiral, Mora initiates conversations and encourages users to openly express their feelings. Throughout these interactions, Mora actively acknowledges and validates users’ thoughts and feelings, creating a safe and supportive conversational environment that facilitates emotional relief.

\textit{Positive thinking} is another strategy~\cite{ref_lightsey}. Users experiencing the Sunday Blues often dwell on frustrations or disappointments from the weekend. Mora addresses this by encouraging users to reflect on their positive weekend experiences, highlighting enjoyable moments or personal achievements to foster feelings of gratitude and contentment. Additionally, Mora nudges users to plan relaxing or entertainment activities for the upcoming weekdays, guiding them to anticipate these pleasurable experiences, thus easing their transition into the workweek.

Finally, Mora integrates the strategy of \textit{problem solving}~\cite{ref_dzurilla}. Returning to work after a restful weekend can lead users to feel overwhelmed by upcoming tasks and responsibilities. Mora addresses this by guiding users to organize their thoughts, prioritize tasks, and formulate clear action plans, enabling them to approach the upcoming workweek with confidence, clarity, and reduced anxiety regarding workload and challenges.

\subsubsection{Video Prototype}
To develop the prototype for Mora, we utilized the \textit{Misty II robot}, an open robotics platform for research and educational purposes~\cite{ref_misty}. We chose Misty II because it can be programmed to display various verbal and non-verbal behaviors aligned with Mora’s intended functions. Moreover, its small size makes it well-suited for home use.

Following Markopoulos’s guidelines for video prototyping~\cite{ref_markopoulos}, we filmed authentic user interactions with Mora within the intended context of use (i.e., Sunday evenings at home), allowing users to immerse themselves deeply in the envisioned experience. Additionally, drawing upon animation techniques~\cite{ref_schulz}, we designed and synchronized Mora’s facial expressions and body movements with its speech to ensure users understand Mora’s emotions and motives, facilitating a more intuitive user-robot interaction.

\begin{figure}
\centering
\includegraphics[width=1\textwidth]{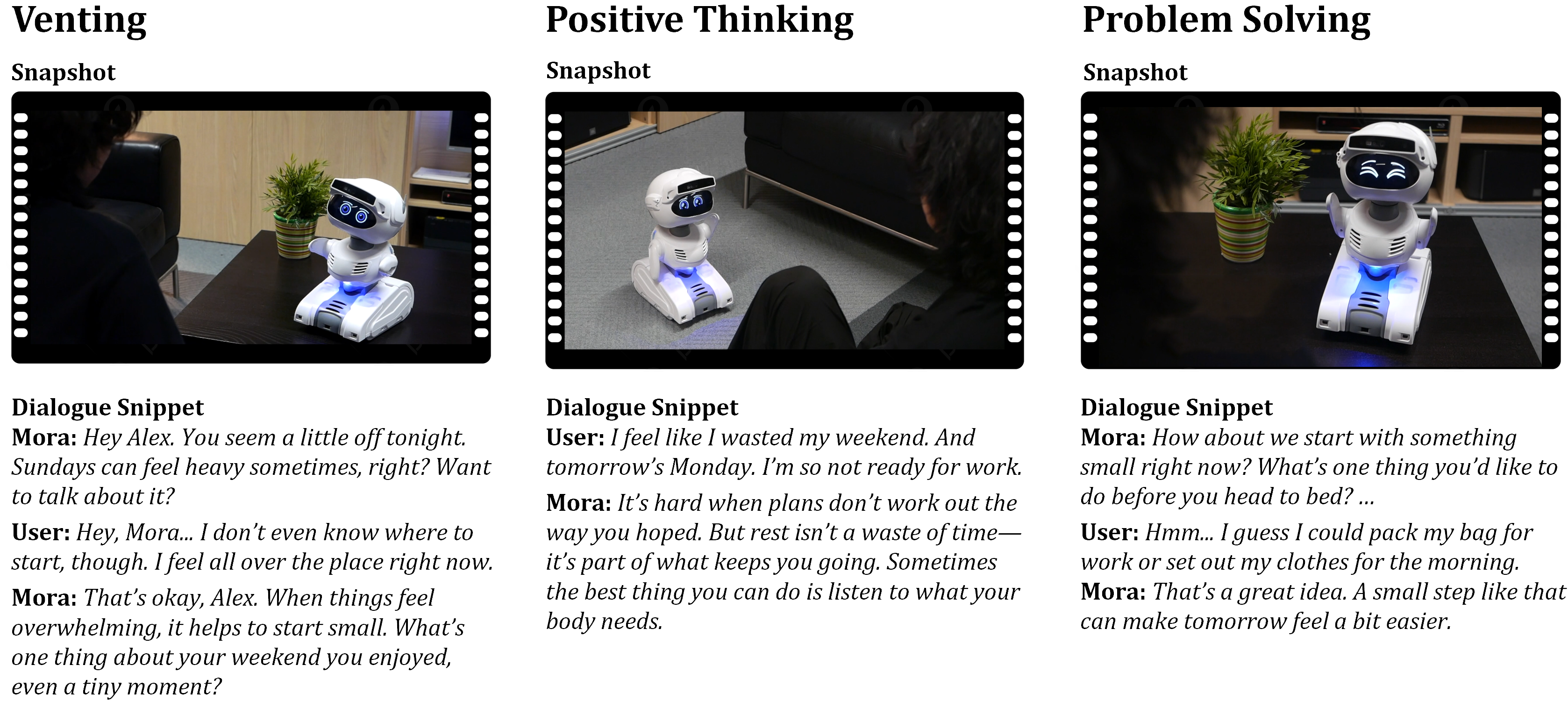}
\caption{Snapshots and dialogue snippets of the video prototype instantiating three mood regulation strategies.} \label{fig1}
\end{figure}

The resulting video introduces Mora and presents two scenarios demonstrating how Mora helps a user to manage the Sunday Blues. In Scenario 1, Mora detects the user's low mood on Sunday evening and initiates a supportive conversation. After discovering the user's disappointment over an unproductive weekend, Mora comforts them by highlighting rest as an essential aspect of productivity. To further reduce the user's negative thinking, Mora suggests preparing their bag for the following day, fostering a sense of preparedness. Similarly, in Scenario 2, Mora proactively engages with the user during a period of anxiety related to upcoming heavy workload. Mora assists with identifying and organizing stressors, planning a manageable task for Monday morning, and scheduling a rewarding self-care activity during the day. To conclude, Mora plays a relaxing playlist to help the user unwind and ease into sleep. Figure~\ref{fig1} presents several snapshots and dialogue snippets of these scenarios, and the full video can be accessed through the provided link (\url{https://vimeo.com/1063895899/31a69169f0}).

\subsection{Participants}
Fifteen participants (aged 24-34; 7 female, 8 male), predominantly researchers from higher education and technology sectors, were recruited through convenience sampling. All participants were employed, typically started their workweek on Monday, and frequently experienced or had previously experienced the Sunday Blues. The sample size was determined following guidance from Hennink and Kaiser~\cite{ref_hennink}, who suggest that 9 to 17 interviews are generally sufficient to reach data saturation. Each participant received a five-euro voucher as compensation, and the study was approved by the Human Research Ethics Committee at the Delft University of Technology (reference number 5088).

\subsection{Co-Constructing Stories}
Based on Buskermolen and Terken’s framework~\cite{ref_buskermolen}, each co-constructing stories session was structured into two phases: (1) \textit{sensitization}, aimed at eliciting participants’ past experiences, and (2) \textit{envisioning}, which encouraged participants to relate these experiences to the design concept and envision future contexts of use.

In the sensitization phase, participants were first asked to describe their typical weekend routines and how they usually felt on Sunday nights. They then watched a short sensitizing video (available at~\url{https://vimeo.com/1064033409/d19fe0b5e2}), which depicted a scenario of someone experiencing the Sunday Blues. To ensure a consistent narrative across phases, this video featured the same character and home environment as the subsequent Mora video prototype. After watching the sensitizing video, participants reflected on their personal experiences related to the scenario. They shared their own encounters with the Sunday Blues and discussed strategies they had developed to manage these feelings.

The envisioning phase started with participants watching the Mora video prototype, which presented a fictional story about how a social robot assisted a user in dealing with the Sunday Blues. After viewing, participants shared their overall impressions of the design concept, pointing out what they liked or disliked in the story. Next, they were asked to imagine themselves as the main character in the video. They described how they would respond in a similar situation, what actions they would take, and any concerns or barriers that might prevent them from using Mora as a coping tool. Finally, participants connected their earlier shared experiences with the Sunday Blues to the design concept, offering suggestions for how it could be improved or adapted to better fit their personal contexts and needs.

Each co-constructing stories session lasted approximately 30 minutes, with all discussions audio-recorded for subsequent analysis.

\subsection{Data Analysis}
All audio recordings were transcribed, and thematic analysis was conducted based on Braun and Clarke’s framework~\cite{ref_braun}. The process involved six steps: (1) familiarization with the data, (2) coding, (3) generating initial themes, (4) reviewing and developing themes, (5) refining themes, and (6) reporting the results. To ensure reliability, two researchers collaborated throughout the process~\cite{ref_clarke}. Familiarization occurred naturally during transcription, so the first author began by independently coding all transcripts and generating initial themes, which produced a preliminary codebook. The second author then independently applied this codebook to the transcripts, critically evaluating the existing codes and themes while suggesting modifications and/or additions. Next, the two researchers discussed discrepancies and refined the categories until they reached a consensus, resulting in a more accurate and comprehensive set of codes and themes. Finally, this refined collection was reviewed and finalized by all authors when reporting the results. Our final coding scheme included 3 themes and 20 codes, presented in Tables~\ref{tab1}, \ref{tab2}, and \ref{tab3}, and further elaborated in the results section.
\section{Results}
\subsection{User Concerns Regarding the Social Robot}
Participants expressed five key concerns regarding Mora as a social robot, specifically focusing on its capabilities and attributes required in the context of mood regulation (Table~\ref{tab1}).

\subsubsection{Limited Humanness in Conversation}
Participants found conversations with Mora unnatural compared to human interactions. They noted that those conversations felt rigid, following a predetermined structure. Additionally, they perceived Mora’s communication style as “sterile” and suggested that Mora should exhibit a more distinctive personality, express more emotions, and occasionally incorporate humor to create a more authentic conversational experience.

\subsubsection{Lack of Genuine Empathy}
Participants acknowledged that Mora’s responses felt supportive and empathetic. They even pointed out that such expressed empathy could be greater than that of humans in certain cases, especially given recent advancements in artificial intelligence. However, deep down inside, participants believed that Mora could not genuinely empathize with them or truly understand their thoughts and feelings, as the robot lacks lived experiences and personal situations similar to their own.

\subsubsection{Limited Social Sensitivity}
Participants raised concerns about Mora’s level of social sensitivity, i.e., its ability to accurately interpret and appropriately respond to users’ social cues and context. They stressed that Mora should be highly sensitive in recognizing when a conversation is unwanted or inappropriate. For instance, Mora should disengage when a user shows little interest, or refrain from initiating a conversation if a user is already occupied, even though early signs of the Sunday Blues are detected.

\subsubsection{Lack of Genuine Rapport with Users}
Participants expressed doubts about Mora’s ability to develop a genuine rapport with users. They were particularly skeptical about whether Mora could intuitively determine when a user desires company or conversation versus when they prefer solitude to process their thoughts and emotions. One participant emphasized this by describing the complex and subtle nature of human rapport: a close friend realizes when to step back and leave someone alone, but also knows precisely when to reapproach to provide comfort and support, even if it temporarily overrides personal autonomy.

\subsubsection{Potential Replaceability}
Some participants questioned the unique value of a social robot like Mora for conversation-based interventions. One participant specifically mentioned that an embodied conversational agent integrated into a smartwatch could be equally effective. Another participant, who regularly practices gratitude journaling via a mobile application, felt Mora’s functions were already met by existing tools, making the presence of a physical robot potentially unnecessary.

\begin{table}
\caption{User concerns regarding the social robot.}
\label{tab1}
\begin{tabulary}{\textwidth}{|L|L|}
\hline
Code & Example participant quote\\
\hline
Limited humanness in conversation & “It’s now a one-way kind of interaction. Like, Mora is asking me questions. I’m replying to Mora. It’s asking me questions again. I think there needs to be more of a dialogue that happens as well as a dialogue with some personality.” (P11) \\ \hline
Lack of genuine empathy & “It’s not a real person or real living who can really accompany you, really understand you, because it couldn’t experience all the things you’re doing.” (P10)\\ \hline
Limited social sensitivity & “If you’re having a conversation with friends on the phone, or if you’re in this important part of your Netflix series or your book, then I would actually be a bit annoyed. So, it would need to know when to talk to you and when not ... so, when you can be interrupted and when not, I guess.” (P4)\\ \hline
Lack of genuine rapport with users & “I think the robot should sense in some way that, ‘Okay, now, enough of venting, I can just leave her some time.’ So, I think that transition needs to happen ... that we talk, and once I know I’m now good and grounded, then I also need to be by myself for some time. I think it would be very similar to talking to a friend.” (P8)\\ \hline
Potential replaceability & “I don’t see that much value that it’s being physical now. I could still imagine, for instance, your watch saying, ‘Hey, you’re stressed, why don’t you want to have this call with whatever?’ And then a kind of avatar or something else could do the same.” (P4)\\
\hline
\end{tabulary}
\end{table}

\subsection{User Concerns Regarding Intervention Effectiveness}
Participants expressed ten main concerns regarding Mora’s effectiveness, highlighting various factors that could impact how it functions in real-life situations (Table~\ref{tab2}).

\subsubsection{Intangibility of Mood}
Participants questioned Mora’s ability to accurately detect the Sunday Blues, as it is a subtle and intangible feeling state that may not manifest through obvious facial or bodily expressions. They also noted that individuals may experience a mix of moods on Sunday night, such as the joy of the weekend alongside anxiety about the upcoming week, making the detection of the Sunday Blues even more challenging.

\subsubsection{Unresolvable Causes of Mood}
Participants mentioned that the Sunday Blues could stem from multiple causes, some of which might be difficult or even impossible to resolve. For example, one participant felt that a heavy workload in the upcoming week was sometimes inevitable, while another believed that the unfavorable social relationships at work contributing to their Sunday Blues could not be effectively addressed.

\subsubsection{Machine Communication Reluctance}
Several participants expressed a general reluctance to communicate with Mora, perceiving it as a “machine” that lacks genuine care and understanding. They felt that conversations with such a “fabricated” entity would induce feelings of awkwardness and discomfort. Additionally, unlike human interactions, where social norms create an obligation to respond, participants perceived communication with Mora as entirely optional, further discouraging engagement.

\subsubsection{Social Withdrawal Tendency}
Participants reported that experiencing Sunday Blues often led to a tendency to withdraw from or reduce engagement in social interactions. They explained that when they were not feeling well, explaining their feelings to others could feel rather taxing and exhausting. Additionally, they expressed a need for solitude during these moments, preferring to process and reflect on their negative thoughts and feelings privately.

\subsubsection{Environmental Intricacies}
Participants pointed out the complexities of their home environments, where the presence of roommates, family members, or pets could influence their interactions with Mora. For instance, one participant anticipated providing inauthentic responses to Mora when a roommate was nearby to maintain privacy. Another participant envisioned that their child might interfere in the conversation, making it difficult to stay engaged.

\subsubsection{Dependence on Human-Robot Bonds}
Participants believed that the depth of their conversations with Mora would largely depend on the relationship they developed with it. They explained that they would only feel comfortable sharing their negative emotions and vulnerabilities if they knew that Mora understood them and their context well. Participants emphasized that building this closer, friendship-like bond with Mora would require a prolonged period of acclimatization and consistently satisfying interactions.

\subsubsection{Potential Counterproductive Effects}
Participants expressed concerns that Mora might unintentionally worsen the Sunday Blues. They explained that individuals are not always conscious of their negative feelings or the underlying reasons. In such cases, Mora’s intervention could inadvertently draw attention to their negativity, as well as Monday’s approach, potentially intensifying their distress. Additionally, participants worried that Mora’s suggestions might be too generic to address their complex personal situations involving life, work, and relationships. If the advice felt repetitive or similar to what they had heard many times before, it could lead to frustration or dissatisfaction rather than relief.

\subsubsection{Potential Loss of Interest}
Participants expressed concerns about sustaining long-term engagement with Mora. They felt that after some weeks, the conversations might become repetitive, leading to boredom and a decline in motivation to continue using Mora.

\subsubsection{Real-Time Intervention Versus Prevention}
Participants noted that Mora’s in-the-moment intervention on Sunday evening or night might be ineffective due to time constraints. For instance, users may not have sufficient time to engage in a suggested mood-regulation activity late at night. Instead, several participants proposed a preventive approach, where Mora could provide support throughout the whole weekend by actively checking in on their feelings and influencing their plans and activities. By ensuring a fulfilling and enjoyable weekend, Mora could help foster a lasting sense of contentment that extends into Sunday night, potentially reducing the intensity of the Sunday Blues.

\subsubsection{Long-Term Mood Resilience}
Some participants noted that while Mora provided comfort and relief, the effects felt temporary. Instead, they emphasized the importance of fostering self-regulation, hoping interactions with Mora would encourage self-reflection and help them recognize effective mood regulation strategies. Ideally, they aimed to develop independent coping mechanisms to manage the Sunday Blues without relying on Mora.

\begin{table}
\caption{User concerns regarding intervention effectiveness.}
\label{tab2}
\begin{tabulary}{\textwidth}{|L|L|}
\hline
Code & Example participant quote\\
\hline
Intangibility of mood & “I think if someone is in a not good mood, they might not have very clear facial expressions, or they just want to lie down on the sofa ... with no bodily language. And if Mora incorrectly detects this and classifies this as the Sunday Blues, then I think the user probably feels kind of annoyed.” (P8)\\ \hline
Unresolvable causes of mood & “[What] if I don’t want to see a colleague in the office? Actually, this is not about work itself, it’s about the relationship in the office.” (P14)\\ \hline
Machine communication reluctance & “I always wonder in real situations, would you have those kinds of conversations with the machine while you’re cognitively knowing this is a machine?” (P3) \\ \hline
Social withdrawal tendency & “Sometimes if you are in the Sunday Blues ... you probably just want to lock your mind and want to self-digest your negative mood. So, in that case, the user probably will gonna refuse to continue the chat with Mora.” (P13)\\ \hline
Environmental intricacies & “For example, you’re living with someone else. You also don’t want to share your thoughts or problems with that person. So, if Mora is there and comes to ask you, you may just give them some wrong answers.” (P5)\\ \hline
Dependence on human-robot bonds & “You would still not, like, start talking about your feelings immediately unless you already had that kind of relationship.” (P11)\\ \hline
Potential counterproductive effects & “Sometimes even though I feel my weekend is wasted, or I feel very reluctant to start my new week ... it’s not that obvious. If somebody just mentioned to me, ‘You look like depressed,’ then it reminds me, ‘Okay, tomorrow is Monday,’ and probably it’s like reinforcing the bad mood.” (P2)\\ \hline
Potential loss of interest & “In the first weeks, you might think it’s a good conversation. But after a few weeks, you [might] realize why there is always a fixed routine, and I can imagine what you are gonna say next. Right? So, I might feel a bit bored.” (P6)\\ \hline
Real-time intervention versus prevention & “I’m thinking that Mora probably can do one step ahead. Like, instead of fixing it after the problem happened, they can prevent this even before this happened. ... Like, on Sunday, [Mora] can just give users some advice, saying, ‘Hey, it seems you didn’t do much today, or it seems you are not going out or not having fun today, I suggest you can have some activities or have some fun, since this is the last day of the weekend.’” (P13)\\ \hline
Long-term mood resilience & “I think a more sustainable way is that the users can learn those strategies, and they can adopt them when they feel depressed during weekends. ... I want to have some technology [that] can help me reflect and can help me do better for the next time after all of this.” (P2)\\
\hline
\end{tabulary}
\end{table}

\subsection{User Concerns Regarding Ethics}
Participants raised five ethical concerns regarding using Mora for mood regulation in home environments (Table~\ref{tab3}).

\subsubsection{Violation of Privacy}
Participants expressed concerns about Mora’s constant monitoring and analysis of their mood states, feeling a sense of surveillance in their own homes. They also worried about data security and privacy, fearing that unauthorized individuals might gain access to their personal lives through Mora’s data, which could include sensitive information about their work, relationships, and other private details.

\subsubsection{Deprivation of Autonomy}
Participants felt a potential loss of control over their own emotional states due to Mora’s constant mood detection and active intervention. They were also concerned that Mora’s frequent suggestions might limit their own reflection and decision making. They described Mora’s approach as “paternalistic,” feeling it repeatedly directed their thoughts and actions, without fully respecting their ability to take responsibility for themselves.

\subsubsection{Overemphasis on Positivity}
Many participants felt that the Sunday Blues was not a severe mood issue, regarding it as a normal part of their weekly rhythms. They expressed concerns that Mora and its approach might place too much emphasis on positivity, potentially leading users to perceive the Sunday Blues as a more serious problem than they previously had. This, in turn, could overshadow the value of accepting negative feelings as a natural part of the human experience.

\subsubsection{Risk of Technology Attachment}
While some participants acknowledged the promising potential of Mora, they also expressed concerns about becoming emotionally attached to it. They envisioned that as Mora became more familiar with their preferences or behaviors, they might feel increasingly inclined to interact with it and rely on its suggestions or interventions. This potential attachment to technology made them feel uneasy and even dreaded.

\subsubsection{Risk of Undermining Human Relationships}
Participants worried that relying on Mora for mood regulation could undermine their relationships with loved ones. They emphasized that there is a positive side of experiencing negative feelings—it creates opportunities to seek social support and strengthen connections with others. If Mora consistently helped manage their moods, they feared it might reduce their motivation to engage with friends and family.

\begin{table}
\caption{User concerns regarding ethics.}
\label{tab3}
\begin{tabulary}{\textwidth}{|L|L|}
\hline
Code & Example participant quote\\
\hline
Violation of privacy & “It’s keeping track of your emotional state ... I would be wary of it, with regards to privacy.” (P7)\\ \hline
Deprivation of autonomy & “When it started giving kind of proper tasks, like, ‘You could do this or this.’ I think in that case I would maybe feel like losing autonomy in a way, like, Mora is starting to make the decision for you.” (P7)\\ \hline
Overemphasis on positivity & “This kind of always being positive could be annoying ... so, it’s like, sometimes when you take this role of being the positive one, then you don’t give the other person space to be negative.” (P10)\\ \hline
Risk of technology attachment & “It feels it would learn [about me] over time ... probably after a while, I could get attached to that, because I’m more curious about it. I don’t know if I would actually want that, because then I would be scared to get too attached.” (P10)\\ \hline
Risk of undermining human relationships & “I would be worried that it would work a little bit, and therefore my need to share my feelings with a friend or with my partner would go down. Therefore, I wouldn’t do that. ... So, I feel like, when you’re not feeling well, it’s kind of an opportunity to share that with other people. And it could be a shame if you don’t do it.” (P10)\\ 
\hline
\end{tabulary}
\end{table}
\section{Discussion}
\subsection{Implications for Designing Social Robots for Mood Regulation}
In this section, we discuss the implications of our findings for designing social robots to support everyday mood regulation. We highlight four design considerations (DC1 to DC4) translated from the study results that HRI practitioners could consider to promote user acceptance and mood intervention effectiveness.

\subsubsection{DC1 - Design Robots with Honest Identities}
Our findings show that participants were clearly aware of the robot’s artificial essence. Although they generally appreciated the social interactions, they still recognized robots as machines lacking genuine understanding and empathy. This aligns with Alač’s argument that while people may engage with robots \textit {as if} they were sentient beings, they ultimately perceive them as material objects~\cite{ref_alac}. This insight leads to an important design takeaway: robots should be designed with \textit{honest identities}. Designers should avoid overstating robots’ emotional capabilities—for example, by claiming that robots truly “understand” or “feel” emotions—as this may lead to user disappointment or distrust. Even though recent advances in large language models have enabled robots to simulate human conversation with remarkable fluency, it is equally important that the interaction remains consistent with the robot’s inherent machine identity to ensure an authentic user experience.

\subsubsection{DC2 - Prioritize Social Bonding}
Our findings suggest that despite participants’ awareness of robots’ inherent machineness, they remained open to forming relationships with robots, implying that establishing a bond is critical for them to feel comfortable opening up about personal mood-related issues. This indicates that, for everyday mood regulation, it is not sufficient for robots to merely deliver professional advice (like a coach); they should also serve as meaningful \textit{social companions}. Without this social bond, users may hesitate to disclose deeper feelings, view robots as easily replaceable, or quickly lose interest. We, therefore, encourage designers of mood-regulating robots to actively integrate relationship-building strategies, such as creating engaging shared activities~\cite{ref_gaggioli}, enabling robots to maintain persistent memories over time~\cite{ref_fox}, or considering the matching between human and robot personalities~\cite{ref_lei}.

\subsubsection{DC3 - Be Mindful of Social Norm Violations}
Our findings indicate that social robot interventions designed for mood regulation were perceived as prone to making social errors. These may include intervening at inappropriate times, incorrectly recognizing user mood states, or intruding on users’ personal spaces. Such errors can significantly undermine users’ perceptions of the robot’s social-affective competence, reduce user willingness to accept interventions, and even provoke social conflict~\cite{ref_tian}. These challenges highlight the importance of designing social robots to be sensitive to both explicit and implicit social norms. This sensitivity may involve enabling robots to proactively solicit users’ willingness to interact, while also accurately interpreting unspoken social signals—such as facial expressions or bodily gestures~\cite{ref_sartori,ref_giuliani}—to assess users’ openness to interaction and thus avoid unintended intrusions. 

\subsubsection{DC4 - Set Intervention Boundaries}
Our findings underscore several ethical challenges. Participants frequently expressed concerns about privacy and loss of control. Some also warned that interacting with robots could distance them from real human relationships—a concern previously raised in HRI research~\cite{ref_friedman}. These challenges must be addressed with care. First, mood-related data collection should be under user control, and robots must clearly communicate what data is being collected and how it is stored~\cite{ref_axelsson}. Second, robots should encourage users to maintain connections with the real world. While designing social robots to support human well-being is a valuable goal, it is equally important to establish clear boundaries for intervention and avoid fostering overdependence.

\subsection{Implications for Designing for Mood Regulation}
Beyond informing the design of social robots, our findings contribute to the broader discourse on designing emotionally supportive technologies within the human-computer interaction (HCI) community~\cite{ref_peng1,ref_peng2,ref_desmet}. We outline four key design considerations (DC5 to DC8) that apply both specifically to social robots and more generally to technologies aimed at supporting mood regulation.

\subsubsection{DC5 - Balance Mood Regulation and Other Fundamental Needs}
Our findings reveal that while interventions may effectively support mood regulation, they may also inadvertently undermine users’ other fundamental needs such as privacy, autonomy, socialization, and personal development, as observed in our case study. To mitigate this risk, we recommend that HCI practitioners proactively investigate and understand these core needs during the early design phase and thoughtfully integrate these considerations into the design process to ensure both intervention effectiveness and user experience.

\subsubsection{DC6 - Tailor Strategies to Mood’s Multifaceted Causes}
Our findings reveal that a single mood-regulation intervention may not effectively address all the underlying causes of a negative mood, especially when these causes stem from broader, more complex problems. To enhance effectiveness, we recommend that HCI practitioners adopt a holistic approach incorporating diverse strategies tailored to different types of stressors. For example, an intervention could facilitate avoidance or suggest direct resolution for identifiable and manageable stressors, while offering relief or distractions for stressors that cannot be easily resolved.

\subsubsection{DC7 - Combine Preventive and Intervening Approaches}
Our findings suggest that real-time interventions may be ineffective for mood regulation due to time and contextual constraints on Sunday evenings or nights. To address this, we found the preventive approach proposed by our participants particularly insightful, where interventions proactively reduce the likelihood of negative moods arising rather than only responding once they occur. We recommend integrating both preventive and intervening strategies: prevention to minimize stressors in advance, and intervention to address or mitigate them when they still arise. This combined approach better aligns with the elusive and long-lasting nature of human mood, ultimately enhancing the effectiveness of mood-regulation interventions.

\subsubsection{DC8 - Respect the Acceptance of Negative Moods}
Our findings indicate that individuals may perceive the Sunday Blues as a normal part of their weekly experience rather than a problem requiring intervention. In such cases, introducing a mood-regulation solution could inadvertently increase awareness of the issue, potentially reframing it as more serious than previously perceived. This shift might lead users to replace their existing, comfortable ways of coping with a new approach focusing on pursuing positivity at all costs. Over time, this could disrupt their natural mood equilibrium, hindering their ability to accept and navigate negative moods in the long term. Hence, we suggest HCI practitioners exercise caution when offering mood-regulation interventions, ensuring they do not overshadow the value of accepting negative feelings as a part of the human experience.

\subsection{Limitations of This Study}
Our exploratory study has several limitations. First, our participants were drawn from a local research community, with most having backgrounds in design and HCI. Their familiarity with emerging technologies may have influenced how they perceived and evaluated Mora. For example, prior HRI studies have shown that individuals with more experience with robots tend to hold less negative attitudes toward them (e.g.,~\cite{ref_rosen}). Future studies should aim to include more experientially diverse populations to capture a broader range of perspectives. Second, participants only viewed a video prototype rather than interacting with a real robot. While the video-based co-constructing story method can be useful for eliciting rich reflections, it may fall short in stimulating the more complex affective responses (e.g., caring and bonding~\cite{ref_jiaxin}) that are associated with physical embodied interactions. Future studies should explore interactions with functional robot prototypes in naturalistic settings to uncover more diverse and ecologically valid user experiences and concerns. Third, our findings are purely qualitative. While we gained a broad overview of user concerns, we did not collect quantitative data to test specific hypotheses. We encourage future HRI research to build on our findings by examining potential statistical relationships between user concern factors, which could help clarify the underlying psychological mechanisms involved in social robots for mood regulation. 
\section{Conclusion}
This article presents a study that explores user concerns regarding social robots designed for mood regulation, specifically focusing on alleviating the “Sunday Blues.” The findings reveal that users expect these robots to possess specific attributes, such as empathy and social sensitivity. Participants also expressed concerns about factors that could affect the effectiveness of robot-based interventions, including reluctance in machine communication and potential counterproductive effects. Additionally, users emphasized ethical considerations regarding the integration of social robots into daily life. Based on these insights, we derive eight design considerations to guide researchers and practitioners in the HRI and HCI communities in developing more effective and ethically sound mood regulation interventions. While the study is limited by its use of video prototyping, it contributes valuable, user-centered empirical knowledge to inform future design practices in social robots for mood regulation, and it opens up opportunities for further exploration of high-fidelity robotic prototypes in real-world settings.
\begin{credits}
\subsubsection{\ackname} This study was funded by the China Scholarship Council (CSC), grant number 202106130007, and the MaGW VICI, grant number 453-16-009, of the Netherlands Organization for Scientific Research (NWO), awarded to Pieter M. A. Desmet.

\subsubsection{\discintname}
The authors have no competing interests to declare that are relevant to the content of this article.
\end{credits}
%
%
%
%

\end{document}